\documentclass[aps,pra,reprint]{revtex4-1}

\usepackage{blindtext}
\usepackage{float}
\usepackage{pict2e}
\usepackage{comment}
\usepackage{mathtools}
\usepackage[utf8]{inputenc}
\usepackage{xcolor}
\usepackage{enumitem}
\usepackage{amsmath,amssymb,amsfonts,bm,braket}
\usepackage{graphicx}
\usepackage[caption=false]{subfig}
\usepackage{hyperref}

\hypersetup{
    colorlinks,
    citecolor=blue,
    linkcolor=blue,
    urlcolor=blue
}

\usepackage[whole]{bxcjkjatype}
\usepackage[normalem]{ulem}
\usepackage{soul}
\definecolor{electricindigo}{rgb}{0.44, 0.0, 1.0}
\definecolor{electricpurple}{rgb}{0.75, 0.0, 1.0}
\definecolor{electricultramarine}{rgb}{0.25, 0.0, 1.0}
\definecolor{americanrose}{rgb}{1.0, 0.01, 0.24}
\newcommand{\kuro}[1]{\textcolor{electricpurple}{{\bf YK: }#1}}

\newcommand{\tr}{\operatorname{Tr}}
\renewcommand{\Re}{\operatorname{Re}}
\renewcommand{\Im}{\operatorname{Im}}

\newcommand{\sz}{\sigma_z}

\begin{document}
\title{Optimal Control to Minimize Dissipation and Fluctuations in Open Quantum Systems Beyond Slow and Rapid Regimes}

\author{Yuki Kurokawa }
\author{Yoshihiko Hasegawa }

\affiliation{
Graduate School of Information Science and Technology,
The University of Tokyo, Bunkyo-ku, 113-8656 Tokyo, Japan}

\begin{abstract}
Optimal control is a central problem in quantum thermodynamics. When minimizing dissipated work and work fluctuations defined via the two-point measurement scheme in open quantum systems, existing approaches largely focus on the rapid- and slow-driving limits, leaving the behavior at intermediate timescales elusive. In this work, by numerically optimizing the driving protocols, we demonstrate that the open quantum systems exhibit distinct optimal structures not captured by the conventional limits. Specifically, in the coherent spin-boson model, we find that the optimal protocol switches discontinuously between distinct locally optimal solutions as the relative weight between dissipation and fluctuations is varied. Furthermore, for a single-level quantum dot coupled to a fermionic reservoir, the optimized protocol develops a characteristic multi-step structure.
\end{abstract}

\maketitle

\section{Introduction}
The optimization of driving protocols that minimize dissipation and fluctuations is a central theme in nonequilibrium stochastic thermodynamics~\cite{SeifertST}. For classical systems such as a Brownian particle, protocols that minimize the mean dissipated work for a fixed total duration are known to exhibit jumps at the beginning and at the end of the protocol~\cite{SchmiedlSeifert2007,ThenEngel2008}. Jump protocols have also been found in minimizing dissipation and work fluctuations in rapidly driven classical systems~\cite{classic_rapid}. In the slow-driving regime, linear-response theory leads to a work fluctuation--dissipation relation in which the dissipated work is proportional to the work variance~\cite{SivakCrooks2012}. At intermediate driving speeds, numerical optimization of smooth protocols reveals richer behavior, including phase-transition-like changes in the optimal protocol structure when trading off mean dissipation against work fluctuations~\cite{phase_transition}.

%Defining work in quantum mechanics is nontrivial because, unlike classical work, it is not represented by a single Hermitian observable in general~\cite{text_book}.\kuro{これも少し変　古典だとエルミート演算子でworkが定義されているような書き方になっている}
Defining work in quantum mechanics is nontrivial because, unlike classical work, there is no single universally established definition for it.
In particular, the naive operator identification $W=H(\tau)-H(0)$ 
%does not, in general, reproduce the correct work statistics and is 
yields work statistics which are generally %incompatible with
incompatible with quantum fluctuation relations except in special cases~\cite{operator_of_work}.
A standard and widely used resolution is the two-point measurement (TPM) scheme~\cite{Kurchan_TPM,Tasaki_TPM,FT,CampisiRMP,EspositoRMP,NoGoWork2017}.
In the TPM scheme, one performs projective energy measurements at the beginning and end of the protocol and defines the stochastic work as the difference of the two outcomes, $w=E_f-E_i$.
If the initial state carries quantum coherence in the energy eigenbasis, the first projective measurement removes it and thus modifies the subsequent dynamics; this is an intrinsic limitation of the TPM definition.
Nevertheless, TPM work satisfies the quantum Jarzynski equality and related fluctuation theorems~\cite{Mukamel2003,Monnai2005,FT,CampisiRMP,EspositoRMP}, and it is the prevailing framework for characterizing work statistics in driven  open quantum systems~\cite{work_moment,slow_drive,rapid_drive}.

In limiting regimes, for open quantum systems governed by Lindblad dynamics~\cite{GKSL,GKS,Lindblad1976,vectorization,Manzano_Lindblad2020}, optimal protocols that minimize dissipated work and work fluctuations within the TPM framework have been obtained analytically.
In the slow-driving limit, where the dynamics remains close to quasistatic,
quantum generalizations of the classical fluctuation--dissipation relation show
that quantum coherence generically prevents the simultaneous minimization of
dissipated work and work variance~\cite{slow_drive,ScandiPerarnau2019}.
In the opposite, rapid-driving limit, protocols that minimize convex combinations of dissipated work and work variance were characterized analytically within a restricted class of protocols consisting of two jumps separated by a constant plateau \cite{rapid_drive}, rather than through a general numerical optimal-control scheme.

At intermediate time scales, Pontryagin-type optimal-control techniques~\cite{PMP} have been applied
to design finite-time driving protocols that minimize dissipation or maximize
performance in Lindblad dynamics~\cite{Cavina2018}. In these approaches the
thermodynamic cost functionals are time-local, depending on the instantaneous
state and control (for example the mean heat or entropy production), and
fluctuations of TPM work are not incorporated explicitly. 
This is primarily because the variance of the TPM work inherently depends on two-time correlation functions, making it non-local in time and thus difficult to formulate as a standard time-local cost functional for optimal control.
Recently, to address a similar non-locality issue in cyclic quantum heat engines, an auxiliary-operator method was introduced to recast such quantities into a time-local form~\cite{erdman_PRR_2023}.
\begin{comment}
In particular, to the best of our knowledge,  there
is currently no numerical framework \kuro{要修正} that treats dissipated work and TPM work
fluctuations on an equal footing for general time-dependent open quantum
systems.
\end{comment}

\begin{comment}
In this work we establish such a framework \kuro{これは要修正} by developing an optimal-control
method that minimizes both dissipation and fluctuations defined within the TPM
scheme for open systems governed by Lindblad dynamics beyond slow and rapid driving regimes. We consider a control parameter $u(t)$ entering both the system
Hamiltonian and the dissipative rates, and introduce an auxiliary operator
$Y(t)$ obeying a linear equation of motion. This auxiliary state recasts the
inherently history-dependent TPM work variance, which is naturally expressed as
a double time integral over two-time correlation functions, into a time-local
single integral over the enlarged state $(\rho(t),Y(t))$. In this way, a linear
combination of dissipation and fluctuations can be written as a standard
time-local cost functional, making it amenable to gradient-based
Pontryagin/GRAPE optimal-control algorithms.
\end{comment}

In this work, we utilize this auxiliary-operator method to investigate optimal driving protocols that minimize both dissipated work and TPM work fluctuations in finite-time processes. 
Consequently, we discover distinct protocol structures at intermediate timescales, such as discontinuous switches between local optima and the emergence of multi-step patterns.
In particular, we consider
\begin{equation}
  J
  = (1-\alpha)\, W_{\mathrm{diss}}
    + \frac{\alpha \beta}{2}\, \sigma_w^2
    + \frac{\kappa}{2}\bigl(u_T - u(T)\bigr)^2,
  \label{eq:cost}
\end{equation}
where $W_{\mathrm{diss}}$ is the dissipated work defined in the TPM scheme,
$\sigma_w^2$ is the TPM work variance obtained through the auxiliary operator $Y(t)$, and the last term
enforces the terminal constraint on the control parameter $u(t)$.
As noted in \cite{erdman_PRR_2023}, because both $\rho(t)$ and $Y(t)$ satisfy linear equations, the gradient of
$J$ with respect to the time-dependent control can be computed efficiently.
This allows us to employ GRAPE-type gradient-based optimal control algorithms
\cite{Khaneja2005} based on Pontryagin's maximum principle~\cite{PMP}
in order to numerically determine the optimal driving protocols without a priori restricting their functional forms.
%in order to obtain numerically optimal continuous protocols.

We demonstrate the resulting framework on two quantum systems.
%examples \kuro{要修正 exampleではない}.
First, for a driven spin--boson model controlled by a time-dependent bias field, we compare the optimized protocols against the rapid-drive approximation: in the incoherent case ($\Delta=0$) they reproduce the endpoint-jump and near-plateau structure at short protocol durations, while for longer durations the intermediate segment is no longer approximately constant, indicating a departure from the rapid-drive description \cite{rapid_drive}.
For the coherent case ($\Delta\neq 0$), the optimized protocols depend strongly on the trade-off parameter $\alpha$, and the observed missing segment in the Pareto front is consistent with a switch between distinct locally optimal protocol families.
Second, for a quantum-dot model coupled to a wide-band metallic lead, the optimization yields multi-step jump protocols with an additional intermediate jump that is not captured by the rapid-drive approximation.

%\section{setup}
\section{preliminaries}
%\sub
%\subsection{Fluctuation and dissipation in the TPM scheme}
%\hase{Sectionタイトルは最初以外はcapitalizeしない。}\kuro{済}

We adopt the 
%two-projective-measurement 
TPM scheme to define work.
%\hase{一度定義した略語は再定義不要}\kuro{済}
Let $T$ be the protocol duration.
In this setting, the mean work is given by the time-integrated power,
\begin{equation}
\label{eq:def_W_sigma}
\langle W\rangle = \int_0^{T}\tr\bigl(\dot H(t)\,\rho(t)\bigr)dt .
\end{equation}
We consider an open quantum system whose dynamics is described by the Lindblad
equation~\cite{GKSL}
\begin{equation}
\label{eq:Lindblad}
\begin{aligned}
\dot\rho(t) =& \mathcal{L}_t\bigl(\rho(t)\bigr) \\
:=& -\,i[H(t),\rho(t)]\\
    &+ \sum_\alpha \gamma_\alpha(t)\left(
      L_\alpha(t) \rho(t) L_\alpha^\dagger(t)
      - \tfrac12\{L_\alpha^\dagger(t) L_\alpha(t),\rho(t)\}
      \right)
\end{aligned}
\end{equation}
%\hase{endequationの後改行を消して、indentをなくす}\kuro{済}
%\kuro{jump operator に時間依存性を入れました}
with jump operators $L_\alpha(t)$ and rates $\gamma_\alpha(t)\ge 0$.
Throughout this work, we set $\hbar=1$ and assume that the system is initially in
the Gibbs state corresponding to $H_0 := H(0)$ at inverse temperature $\beta$,
\begin{equation}
\label{eq:initialGibbs}
  \rho(0) = \pi_0
  := \frac{e^{-\beta H_0}}{Z(H_0)}, \qquad
  Z(H) = \tr\bigl(e^{-\beta H}\bigr),
\end{equation}
where $Z(H)$ is the partition function.
The dissipated work is then defined as
\begin{equation}
\label{eq:dissipation}
W_{\mathrm{diss}} := \langle W\rangle - \Delta F,
\end{equation}
where $F(H) = -\beta^{-1} \log Z(H)$ and
$\Delta F = F(H_T) - F(H_0)$.

With these conventions, the work fluctuations can be written as
\cite{work_moment,slow_drive,rapid_drive}
\begin{equation}
\label{eq:fluctuation}
\sigma_w^2
= 2\int_0^T dt_1\int_0^{t_1}dt_2
\tr\left\{\dot H(t_1)
\overleftarrow{\mathcal{P}}(t_1,t_2)\bigl[S_{\rho(t_2)}(\dot H(t_2))\bigr]\right\},
\end{equation}
%\hase{$P$にmathcalが抜けている}\kuro{済}
%\hase{ここのTは$\tau$と同じ？}\kuro{式6のTは計算時間、式9の$\tau$は２重積分の内側の積分変数です 式9の$\tau$は式6の$t_2$と同じではあります。そこを修正したほうがいいと言うことでしょうか？}
where
\begin{align}
\overleftarrow{\mathcal P}(t_1,t_2)
&:= \overleftarrow{\mathcal T}\exp\Bigl(\int_{t_2}^{t_1} d\nu\,\mathcal L_\nu\Bigr),\label{eq:propagator}\\
S_\rho(O) &:= \tfrac{1}{2}\{\rho,\,\Delta_\rho O\}_+,
\quad
\Delta_\rho O := O - \tr(O\rho),
\end{align}
and $\{\cdot,\cdot\}_+$ denotes the anticommutator.
Here $\overleftarrow{\mathcal{T}}$ is the time-ordering operator and $\overleftarrow{\mathcal{P}}(t_1,t_2)$ is the time-ordered propagator generated by
$\mathcal{L}_t$, such that
$\rho(t_1) = \overleftarrow{\mathcal{P}}(t_1,t_2)\rho(t_2)$ for $t_1\ge t_2$.

As shown in appendix~\ref{sec:methods}, by introducing
\begin{equation}
\label{eq:Y_def}
Y(t) := \int_0^{t}
\overleftarrow{\mathcal P}(t,\tau)\bigl[S_{\rho(\tau)}(\dot H(\tau))\bigr]\,d\tau,
\end{equation}
$\sigma_w^2$ can be written as 
\begin{equation}
\label{eq:sigma_single_integral}
\sigma_w^2 = 2\int_0^T \tr\bigl(\dot H(t)\,Y(t)\bigr)\,dt .
\end{equation}
%\hase{カンマをピリオドに}
%As shown in Appendix~\ref{sec:Y(t)}, 
%\kuro{appendix F を A に吸収させる}
$Y(t)$ obeys the following time-local equation of motion\cite{erdman_PRR_2023}:
\begin{equation}
\label{eq:Y_ode}
\dot Y(t) = \mathcal L_t\bigl[Y(t)\bigr] + S_{\rho(t)}(\dot H(t)),\qquad
Y(0)=0 .
\end{equation}
This formulation transforms the work variance into a time-local running cost, making it amenable to gradient-based optimization techniques such as the GRAPE algorithm. In the following, we apply this framework to perform numerical optimization for two representative systems: a two-level spin-boson model and a single-level quantum dot.

\section{numerical experiment}\label{sec:numerical-example}
\begin{comment}
In this section, we present the optimized protocols for the spin-boson and quantum-dot models. For the spin-boson model, the optimization produced jump-like protocols, and for $\Delta \neq 0$ the Pareto front showed an apparent missing segment, suggesting a switch between different locally optimal protocol families as $\alpha$ was varied. For the quantum-dot model, the optimization produced multi-step protocols with an additional intermediate jump.

\kuro{ここの部分をmain result にするような書き方にしないといけない}
\kuro{特殊なプロトコルが得られたのがリザルトになるような書き方}
\kuro{数式のタイプセットが所々おかしい}
\end{comment}

\begin{comment}
In this section, by numerically optimizing the driving protocols, we demonstrate that the spin-boson and quantum-dot models exhibit optimal structures not captured by the conventional slow- and rapid-driving limits. Specifically, we show that varying the relative weight between dissipation and fluctuations induces discontinuous switches in the spin-boson model, while a distinct multi-step structure emerges in the single-level quantum dot.
\end{comment}

\subsection{Two level spin-boson model}\label{subsec:spin-boson}

%As a numerical example,
%\kuro{resultであってnumerical example ではない}
We consider the following single spin system in a bosonic reservoir \cite{GKSL,Leggett_RMP1987,adiabatic_GKSL}. %\kuro{\cite{GKSL,adiabatic GKSL}}\hase{引用}. 
The dynamics is described by the Lindblad equation
\begin{align}
\dot{\rho}(t) &= \mathcal{L}_{t}(\rho(t)) \notag\\
&= -i [ H(t),\rho(t) ] \notag\\
&\quad + \gamma_{\downarrow}(E(t)) \left( \sigma_{-}(t)\rho(t)\sigma_{+}(t) - \frac{1}{2}\{\sigma_{+}(t)\sigma_{-}(t),\rho(t)\} \right) \notag\\
&\quad + \gamma_{\uparrow}(E(t)) \left( \sigma_{+}(t)\rho(t)\sigma_{-}(t) - \frac{1}{2}\{\sigma_{-}(t)\sigma_{+}(t),\rho(t)\} \right),
\label{eq:lindblad_generic}
\end{align}

%where $H(t)=u(t)\sigma_z$ with $u(t) \ge 0 , \sigma_+ = \ket{e}\bra{g} , \sigma_- = \ket{g}\bra{e}, \sigma_z= \ket{e}\bra{e}-\ket{g}\bra{g}  $ .
\begin{comment}
\kuro{ ,where $H(t)= \sigma_z u(t) + \Delta \sigma_x.$
Here $ \ket{e(t)}$ and $\ket{g(t)}$ denote the instantaneous excited and ground eigenstates of $H(t)$, forming an orthonormal basis. 
We also define $\sigma_+(t) = \ket{e(t)}\bra{g(t)}$, $\sigma_-(t) = \ket{g(t)}\bra{e(t)}$, and
$\sigma_z= \ket{e(t)}\bra{e(t)}-\ket{g(t)}\bra{g(t)}$.}
\end{comment}
where $H(t)=u(t)\sigma_z+\Delta\sigma_x$.
We work in a fixed basis $\{\ket{g},\ket{e}\}$ and define
$\sigma_z:=\ket{e}\bra{e}-\ket{g}\bra{g}$,
$\sigma_x:=\ket{e}\bra{g}+\ket{g}\bra{e}$.
Let $\ket{e(t)}$ and $\ket{g(t)}$ denote the instantaneous excited and ground eigenstates of $H(t)$, forming an orthonormal basis, and define the corresponding jump operators
$\sigma_+(t):=\ket{e(t)}\bra{g(t)}$ and $\sigma_-(t):=\ket{g(t)}\bra{e(t)}$.
For the spin-boson model, the transition rates are given by
\begin{comment}
\begin{align}
\gamma_{\downarrow}(u) &= \gamma(u)(P(u)+1), \notag\\
\gamma_{\uparrow}(u) &= \gamma(u)P(u),
\label{eq:sb_rates}
\end{align}
\end{comment}
\begin{align}
\gamma_{\downarrow}(E) &= \gamma(E)(P(E)+1), \notag\\
\gamma_{\uparrow}(E) &= \gamma(E)P(E),
\label{eq:sb_rates}
\end{align}
where $\gamma(E)=k E^{3}$ (we set $k=1$) and $P(E)=(e^{2\beta E}-1)^{-1}$ is the Bose-Einstein distribution, where $E(t):=\sqrt{u(t)^2+\Delta^2}$ is the energy gap and the rotation angle is defined by
\begin{equation}
\cos\theta=\frac{u}{E},\qquad \sin\theta=\frac{\Delta}{E}.
\label{eq:theta_def}
\end{equation}
Our aim is to optimize the protocol $u(t)$ in order to minimize dissipation and fluctuation between $u(0)=0$ and $u(T)=1$ at $\beta=1$.
In the numerical implementation, we discretize the total duration $T$ into $N = 1000 \times T$ time steps and set the penalty parameter for enforcing the boundary conditions to $\kappa = 10$. The details of calculation methods and parameters are explained in appendix~\ref{appendix:spin-boson}.

\begin{figure}[t]
  \centering
  \includegraphics[width=\linewidth]{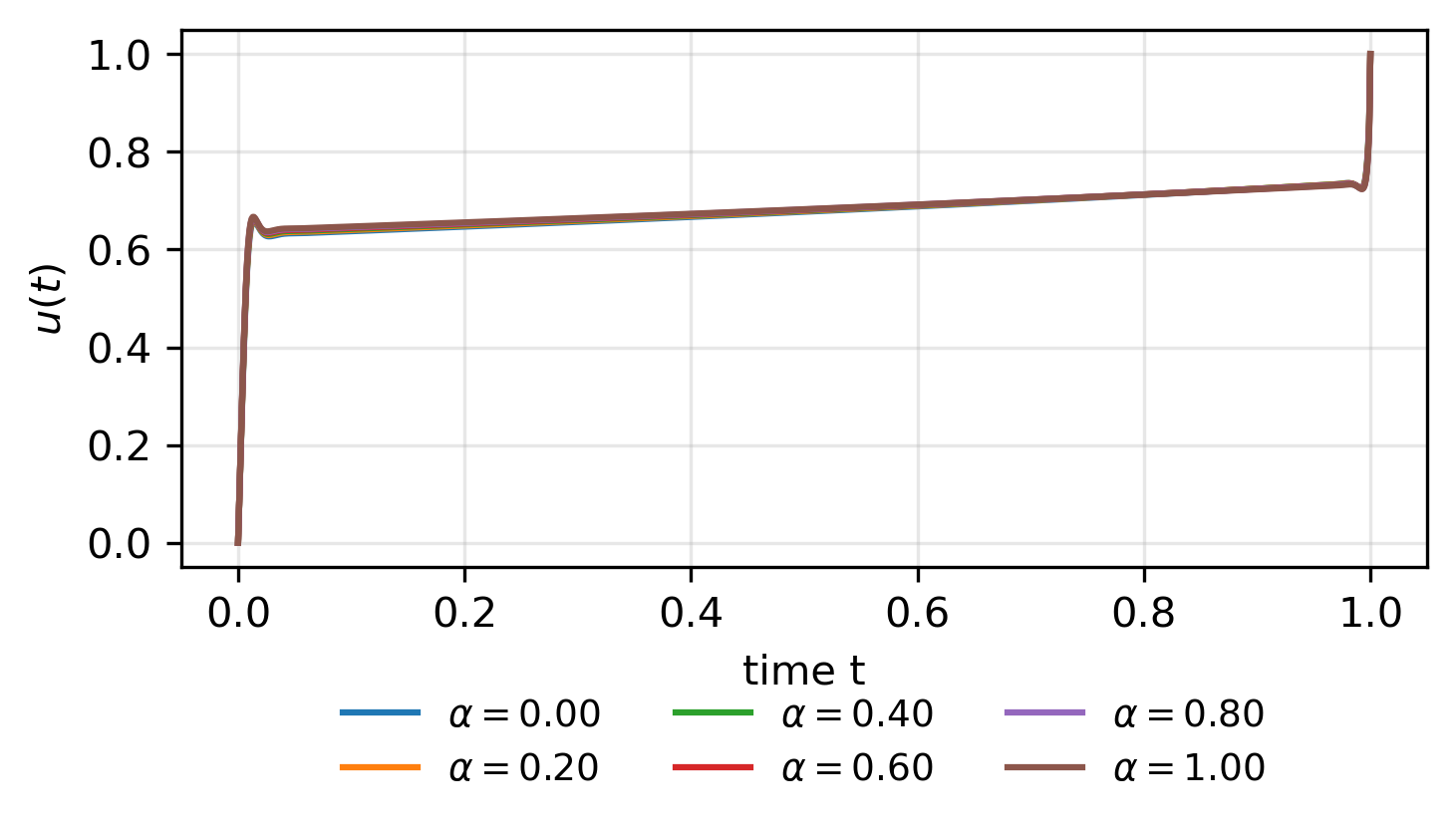}
  \caption{Optimized control field
  $u(t)$ obtained by the GRAPE algorithm when $\Delta=0$ for several values of the weight
  parameter $\alpha$ in the cost functional, as indicated in the legend,
  for protocol duration $T=1$}
  \label{fig:Delta=0_T=1}
\end{figure}

\begin{figure}[t]
  \centering
  \includegraphics[width=\linewidth]{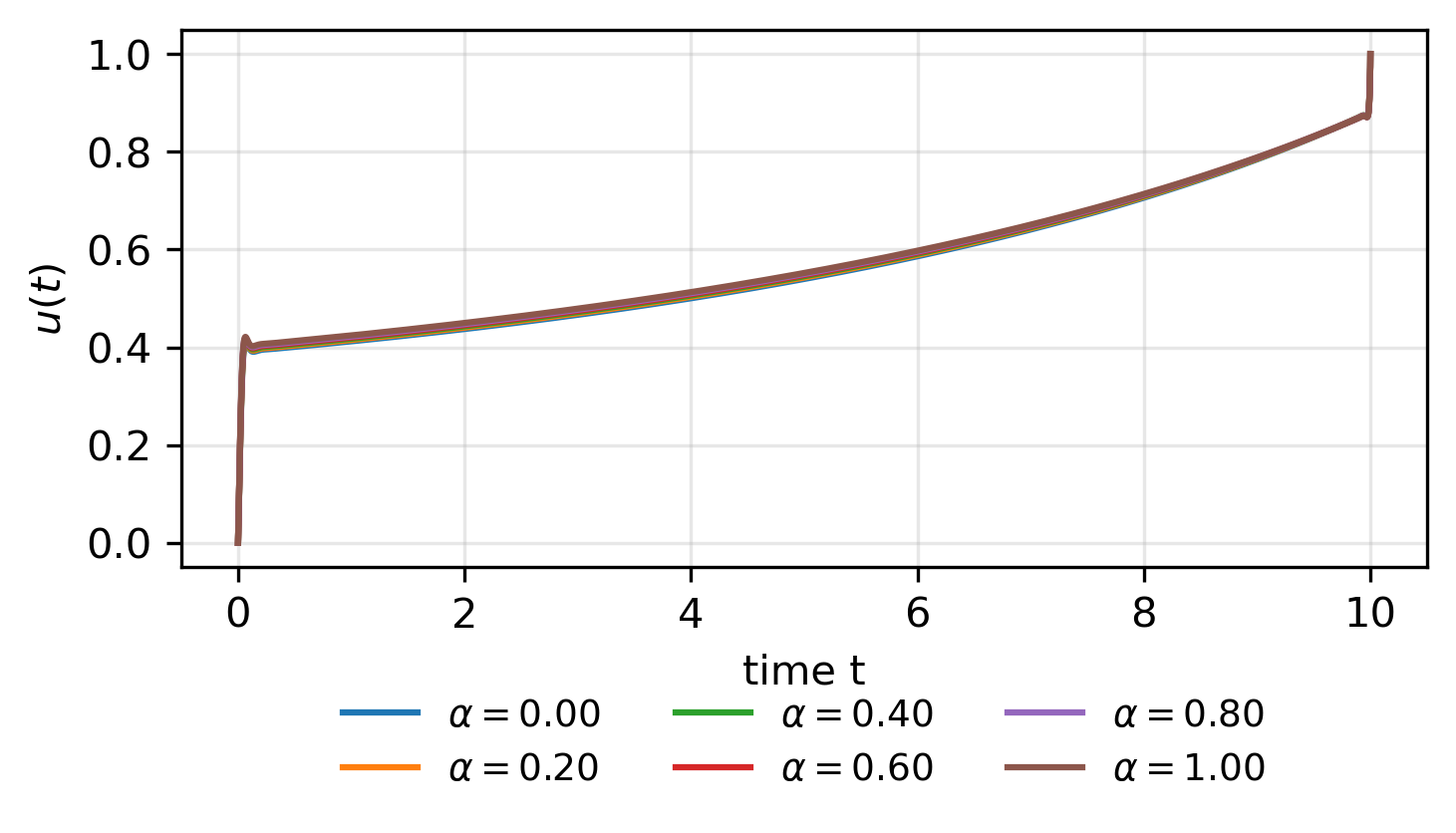}
  \caption{Optimized control field
  $u(t)$ obtained by the GRAPE algorithm when $\Delta=0$ for several values of the weight
  parameter $\alpha$ in the cost functional, as indicated in the legend,
  for protocol duration $T=10$}
  \label{fig:Delta=0_T=10}
\end{figure}

\begin{figure}[t]
  \centering
  \includegraphics[width=\linewidth]{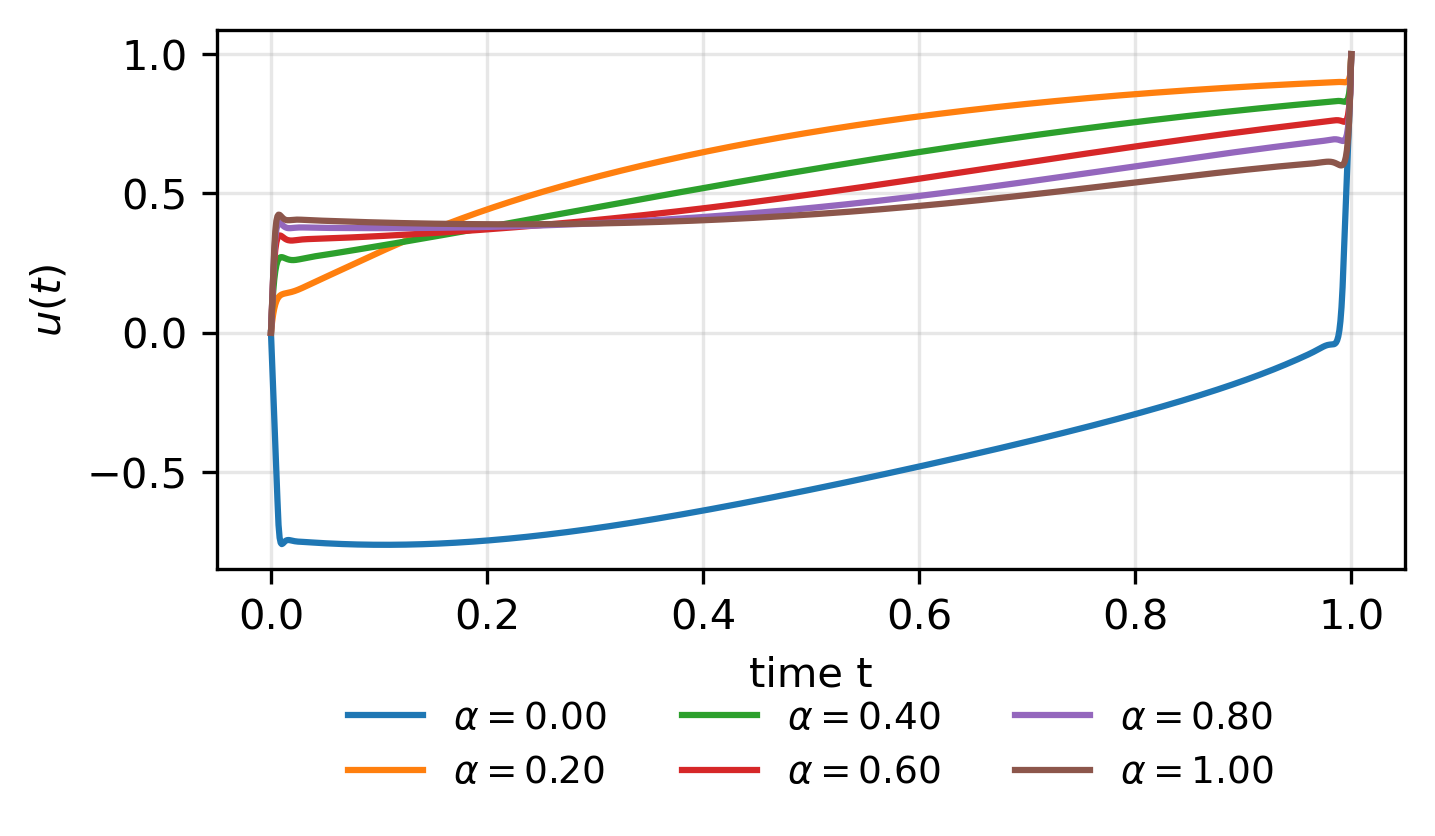}
  \caption{Optimized control field
  $u(t)$ obtained by the GRAPE algorithm when $\Delta=1$ for several values of the weight
  parameter $\alpha$ in the cost functional, as indicated in the legend,
  for protocol duration $T=1$}
  \label{fig:Delta=1_T=1}
\end{figure}

\begin{figure}[t]
  \centering
  \includegraphics[width=\linewidth]{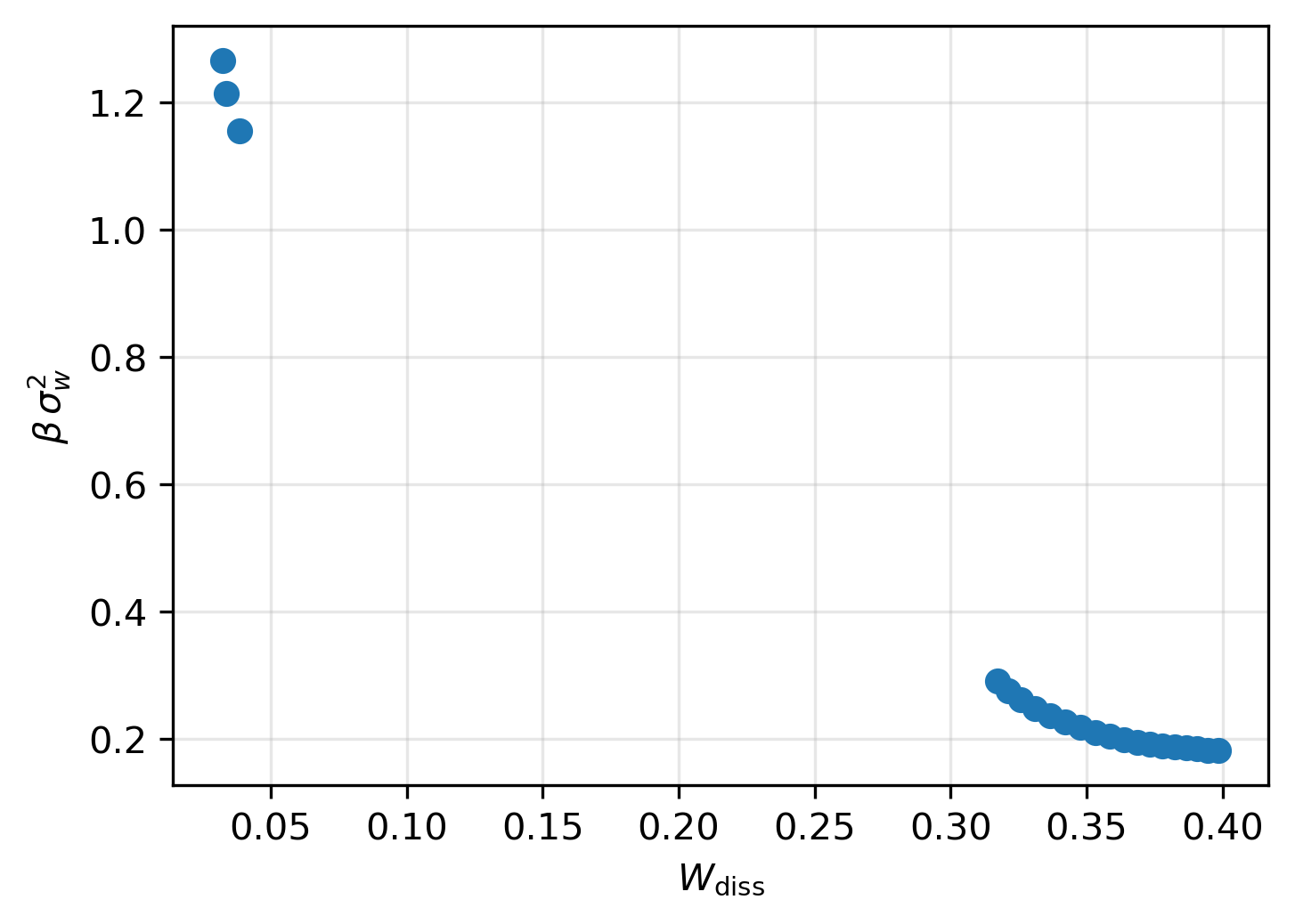}
  \caption{Pareto front for the spin-boson model at $T=1, \Delta=1$ in the $(W_{\rm diss},\,\beta\sigma_w^2)$ plane.
Each marker corresponds to an optimized protocol (obtained by GRAPE initialized from a linear ramp) for a given $\alpha$, where $\alpha$ ranges from $0$ to $1$ in steps of $0.05$.}
  \label{fig:paretp_front_Delta=1_T=1}
\end{figure}

\begin{comment}
For example, when $\Delta=0 ,\quad H(t)=u\sigma_z$, as shown in Fig.~\ref{fig:Delta=0_T=1}, the optimal protocol develops jumps at both endpoints without imposing it beforehand, starting from liner initial guess.
Moreover, as discussed in Ref.~\cite{rapid drive} and Sec.~\ref{sec:rapid drive} , the rapid-driving limit is known to yield protocols with endpoint jumps and constant between them.
In this case, the protocol depends only weakly on $\alpha$, suggesting that dissipation and fluctuations are optimized simultaneously.
\end{comment}
%\kuro{ここまでappendixへ移動}
%\kuro{物理的なパラメータ設定だけ述べて結果に移る、計算に関してはappendixへ}

For the short-duration case ($T=1$) with $\Delta=0$ (i.e., $H(t)=u(t)\sigma_z$), Fig.~\ref{fig:Delta=0_T=1} shows that the optimal protocol develops sharp jumps at both endpoints without imposing such a structure \emph{a priori}, starting from a linear initial guess.
Moreover, as discussed in Ref.~\cite{rapid_drive} and Sec.~\ref{sec:rapid drive}, the rapid-driving limit is known to yield protocols with endpoint jumps and an approximately constant value in between.
In this case, the protocol depends only weakly on $\alpha$, suggesting that dissipation and fluctuations can be optimized simultaneously.

\begin{comment}
As in Fig~\ref{fig:Delta=0_T=10}, for longer time scale $T=10$, although there are jumps at the begging and the end, the protocol between them deviate from constant protocol, which means the deviation from rapid drive approximation.
\end{comment}
As shown in Fig.~\ref{fig:Delta=0_T=10}, for the longer protocol duration $T=10$, the optimal protocol still exhibits jumps at the beginning and the end; however, the intermediate segment is no longer approximately constant, indicating a departure from the rapid-driving approximation.

\begin{comment}
For another example $\Delta=1 ,\quad H(t)=u(t)\sigma_z + \sigma_x$, as shown in Fig.~\ref{fig:Delta=1_T=1}, the optimal protocol minimizing dissipation and fluctuations differs. 
The Pareto front shown in Fig.~\ref{fig:paretp_front_Delta=1_T=1} shows that the trade-off between the objectives differs qualitatively, and part of the Pareto front is missing.
\end{comment}

As another experimental setting, for $\Delta=1$ (i.e., $H(t)=u(t)\sigma_z+\sigma_x$), Fig.~\ref{fig:Delta=1_T=1} shows that the optimal protocol depends strongly on the weight parameter $\alpha$ and differs qualitatively from the $\Delta=0$ case.
In particular, the dissipation-minimizing protocol (small $\alpha$) and the fluctuation-minimizing protocol (large $\alpha$) exhibit different temporal structures rather than being related by a small deformation.
The Pareto front in Fig.~\ref{fig:paretp_front_Delta=1_T=1} indicates a qualitatively different trade-off between dissipation and fluctuations, and we observe a missing segment in the Pareto front.
A plausible explanation is that the optimization switches between two distinct local minima corresponding to different protocol families as $\alpha$ is varied, which can leave an apparent gap when the intermediate trade-off solutions are not selected.

\subsection{Quantum-dot model}
%\kuro{example と言わない方がいい}
As a second numerical setting, we consider a single-level quantum dot weakly
coupled to a fermionic reservoir.  We describe the empty and singly occupied
dot states by the pseudo-spin basis $\{\ket{g},\ket{e}\}$, with
$\sz\ket{g}=-\ket{g}$ and $\sz\ket{e}=+\ket{e}$ as shown in the Appendix.~\ref{sec:qd_fock_to_tls}. 
The system Hamiltonian is again taken as
\begin{equation}
\label{eq:qdot_hamiltonian}
 H(t) = u(t)\,\sz ,
\end{equation}
where $u(t)$ now denotes half the level energy measured from the chemical
potential of the lead, so that the single-particle excitation energy is
$\varepsilon(t)=2u(t)$.  In contrast to the spin--boson model,
$u(t)$ is allowed to take both positive and negative values, corresponding to
moving the dot level above and below the Fermi energy.

In the weak-coupling and Markovian limit, and for a wide-band metallic lead at inverse temperature $\beta$ with chemical potential set to zero, the dot dynamics is described by a fermionic Lindblad equation \cite{Harbola2006}.
Since the control Hamiltonian is the same as in the $\Delta=0$ spin--boson case, 
\begin{comment}
$H(t)=u(t)\sigma_z$, the corresponding control operator $G=\partial_u H=\sigma_z$ is unchanged.
Therefore the auxiliary vector field $\bm{s}(\bm{x})$ and its Jacobian $J(\bm{x})$ are identical to Eqs.~\eqref{eq:s_vector} and \eqref{eq:j_matrix}.
The 
\end{comment}
the modification in the quantum-dot model enters through the dissipative transition rates, which are determined by the Fermi--Dirac distribution $f(u)=(1+e^{2\beta u})^{-1}$ and a constant tunneling rate $\Gamma$:
\begin{align}
\gamma_{\uparrow}(u) &= \Gamma f(u), \notag\\
\gamma_{\downarrow}(u) &= \Gamma\bigl[1-f(u)\bigr], \notag\\
\gamma_{\Sigma}(u) &:= \gamma_{\uparrow}(u)+\gamma_{\downarrow}(u)=\Gamma,
\end{align}
where we set $\Gamma=1$.
\begin{comment}
Accordingly, the vectorized generator $A(u)$ has the same matrix structure as in Eq.~\eqref{eq:matrix_A} (at $\Delta=0$), with the rate-dependent entries given by $\gamma_{\uparrow,\downarrow}(u)$ above.
Consequently, 
\end{comment}
Then, we can apply the same GRAPE optimization scheme as in the spin--boson case,
imposing fixed boundary values $u(0)=u_0=2$ and $u(T)=u_T=-2$ 
and optimizing the
protocol. %$v(t)=\dot u(t)$.
\begin{comment}
Firstly we optimize protocol at $\beta=1$, and obtain the $u(t)$ as in Fig.\ref{fig:qd_alpha}.

Then we change $\beta$ at $\alpha=1$,
\end{comment}

\begin{figure}
  \centering
  \includegraphics[width=0.48\textwidth]{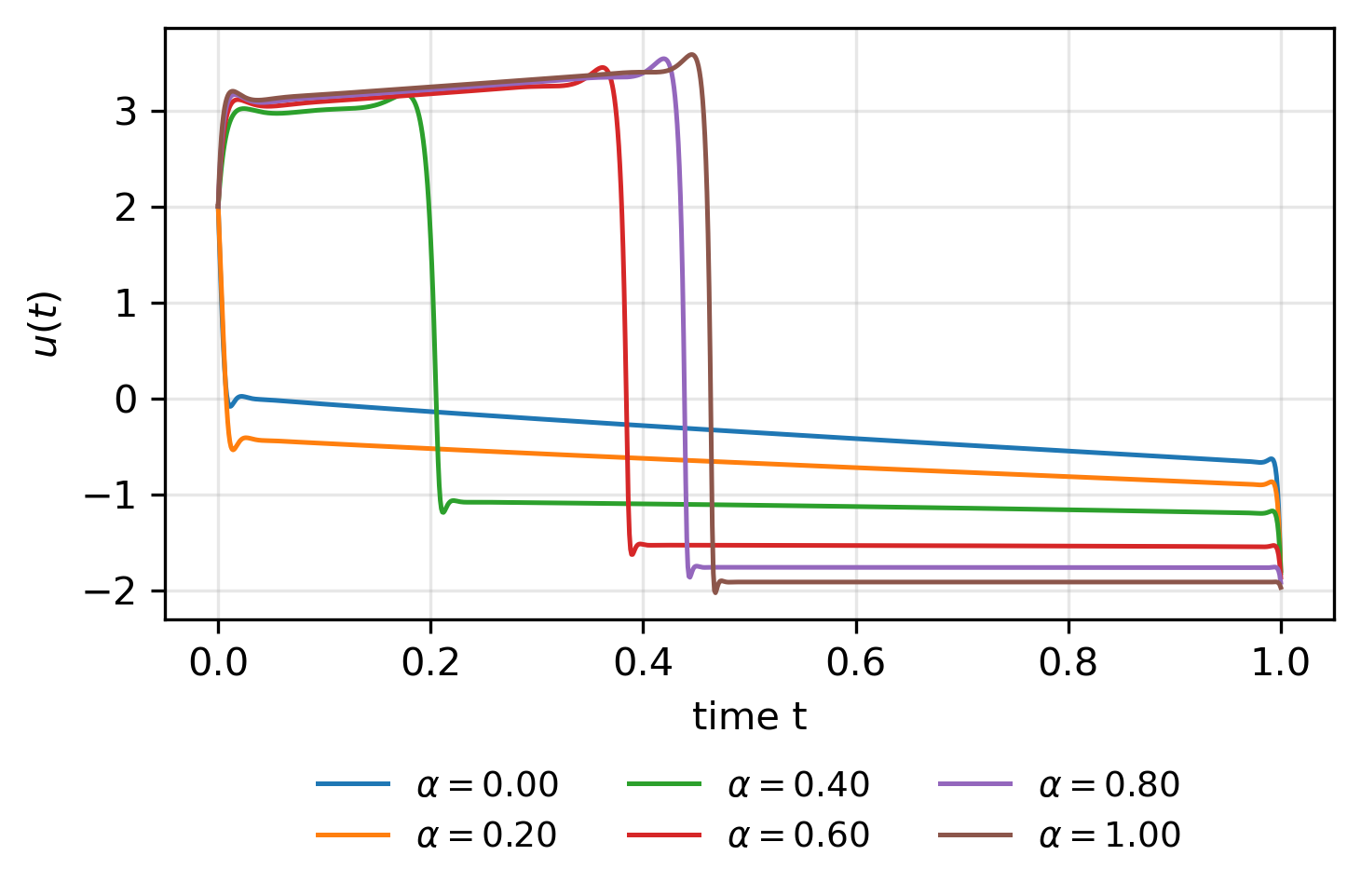}
  \caption{
  Optimized control field u(t) obtained by the GRAPE algorithm for the quantum-dot model at $\beta=1$ for several values of the weight parameter $\alpha$ in the cost functional, as indicated in the legend, for protocol duration T=1.
  %Optimized control protocols $u(t)$ for the quantum-dot model obtained by GRAPE at fixed inverse temperature $\beta=1$ for different values of $\alpha$, with the optimization initialized from a linear ramp.
  }
  \label{fig:qd_alpha}
\end{figure}

\begin{figure}
  \centering
  \includegraphics[width=0.48\textwidth]{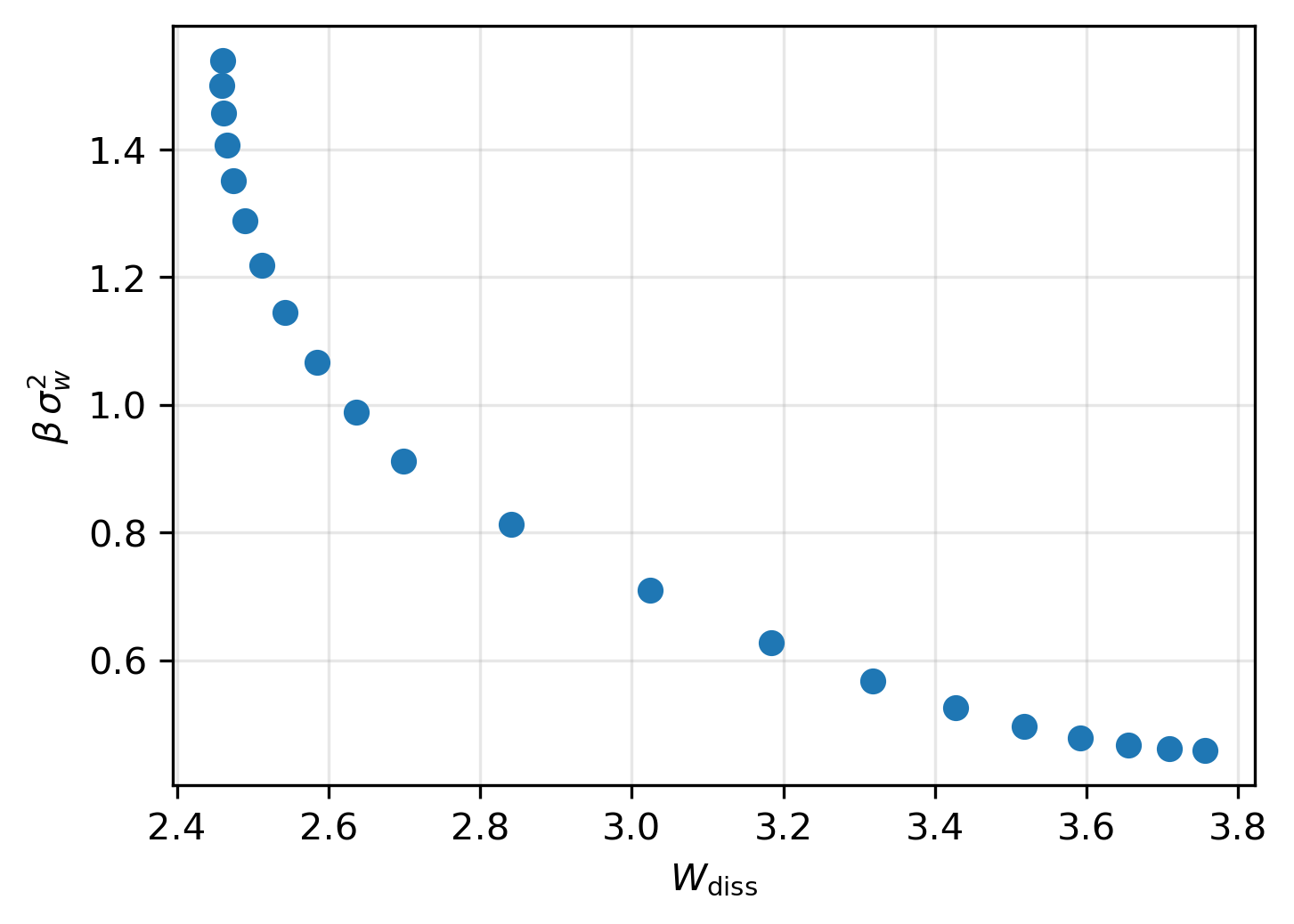}
  \caption{Pareto front for the quantum-dot model at $\beta=1$ in the $(W_{\rm diss},\,\beta\sigma_w^2)$ plane.
Each marker corresponds to an optimized protocol (obtained by GRAPE initialized from a linear ramp) for a given $\alpha$, where $\alpha$ ranges from $0$ to $1$ in steps of $0.05$.
  }
  \label{fig:qd_pareto}
\end{figure}

\begin{figure}
  \centering
  \includegraphics[width=0.48\textwidth]{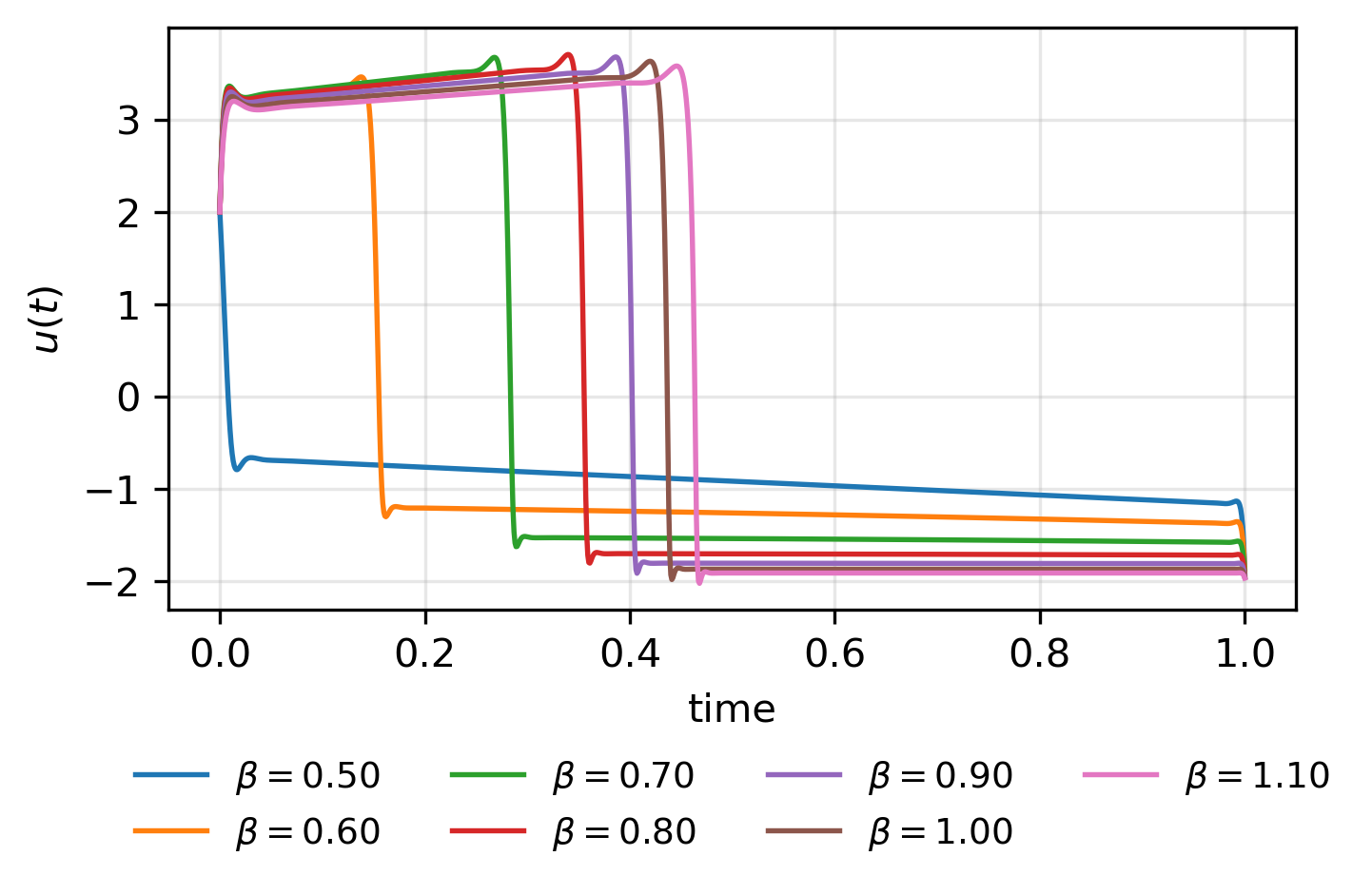}
  \caption{ Optimized control protocols $u(t)$ for the quantum-dot model obtained by GRAPE at $\alpha=1$ for different inverse temperatures $\beta$, for protocol duration $T=1$.
  }
  
  \label{fig:qd_beta}
\end{figure}

Firstly, we fix the inverse temperature at $\beta=1$ and optimize the control protocol %$v(t)=\dot u(t)$ 
for several values of the trade-off parameter $\alpha$.
The resulting optimal controls $u(t)$ are shown in Fig.~\ref{fig:qd_alpha}.
For small $\alpha$, where the cost functional mainly penalizes dissipation, the protocols are shaped primarily to reduce $W_{\mathrm{diss}}$ and are consistent with the rapid-driving solution.
As $\alpha$ is increased and more weight is placed on suppressing work fluctuations, the optimal solution changes qualitatively; beyond a certain range of $\alpha$, $u(t)$ develops an additional intermediate jump and takes on a distinct protocol structure.
%As $\alpha$ is increased and more weight is placed on suppressing work fluctuations, the optimal solution changes qualitatively; beyond a certain range of $\alpha$, the structure of $u(t)$ は中間ジャンプを持つ特殊なプロトコルが生じている。
%is rearranged in a manner reminiscent of the phase-transition--like changes of optimal protocols discussed in Ref.~\cite{phase transition}.
Figure~\ref{fig:qd_pareto} shows the corresponding Pareto front at $\beta=1$, i.e., the set of achievable pairs $(W_{\mathrm{diss}},\,\beta\sigma_w^2)$ obtained from the optimized protocols as $\alpha$ is varied.

Then, we fix $\alpha=1$ (fluctuation minimization) and vary the inverse temperature $\beta$.
Figure~\ref{fig:qd_beta} shows the corresponding optimized protocols $u(t)$ for several values of $\beta$, obtained by GRAPE initialized from a linear ramp.
For small $\beta$ (high temperature), the protocols closely follow the rapid-drive solution consisting of two endpoint jumps separated by a single plateau.
As $\beta$ is increased, the numerically optimal protocol develops an additional intermediate jump and acquires a three-step structure.
By combining the auxiliary-operator method with the GRAPE algorithm, we obtain optimal protocols with a multi-step jump structure that are not captured by either the slow-driving or the rapid-driving approximation.
\begin{comment}
This trend is consistent with the fact that at high temperatures the Fermi--Dirac occupation $f(u)=(1+e^{2\beta u})^{-1}$ depends only weakly on $u$, whereas at low temperatures $f(u)$ becomes increasingly step-like, so even modest changes in $u$ drive the dot close to being fully occupied or empty and the rapid-drive approximation ceases to be valid.
\end{comment}
This behavior can be understood as follows. In the rapid-driving approximation, the state is expanded around the initial equilibrium state and therefore remains close to it throughout the protocol. As $\beta$ increases, however, the Fermi--Dirac occupation $f(u)=(1+e^{2\beta u})^{-1}$ becomes increasingly step-like, so that even modest changes in $u$ induce substantial changes in the dot occupancy. In this regime, the state no longer remains close to its initial value, and the short-time description underlying the rapid-driving approximation breaks down, making an additional intermediate jump favorable.

\section{Conclusion}
In this work, we have investigated the optimal control protocols that minimize both dissipated work and work fluctuations in open quantum systems, specifically focusing on the intermediate-time regime beyond the conventional slow- and rapid-driving limits. By utilizing the auxiliary operator method to recast the inherently non-local TPM work variance into a time-local form, we cast the simultaneous minimization of dissipation and fluctuations into an optimal control framework. This approach enabled the efficient application of a gradient-based optimization algorithm (GRAPE) to find the optimal driving protocols without relying on approximations.

By applying this framework to a two-level spin-boson model, we revealed that the presence of quantum coherence fundamentally alters the optimal control landscape. For the incoherent case ($\Delta=0$), the optimal protocols continuously deform between the dissipation-minimizing and fluctuation-minimizing solutions. In contrast, for the coherent case ($\Delta \neq 0$), we found that the optimal protocol switches discontinuously between distinct local optima as the relative weight between the two objectives is varied. This switching behavior manifests as a distinct gap in the Pareto front, highlighting a highly non-trivial trade-off relation induced by quantum coherence.

\begin{comment}
Furthermore, our analysis of a single-level quantum dot coupled to a fermionic reservoir demonstrated the emergence of a characteristic multi-step structure in the optimal protocols. 
このプロトコルはjump を含むためslow ではなく、またrapid driveのように両端のみのジャンプでなく中間にジャンプを含んでおり、これまでに観察されたプロトコルと定性的に違うものが得られた。
We clarified that
clarify は言い過ぎ　推測できる　程度
this intermediate jumping behavior originates from the sharp step-like nature of the Fermi-Dirac distribution at low temperatures, where a small change in the energy level causes a drastic change in the system's occupancy. This result clearly shows that the simple jump-and-hold paradigm breaks down, and intermediate steps are actively required to suppress severe work fluctuations while managing dissipation.
\end{comment}

Furthermore, our analysis of a single-level quantum dot coupled to a fermionic reservoir demonstrated the emergence of a multi-step structure in the optimal protocols. Since these protocols contain intermediate jumps, they differ qualitatively from continuous protocols in the slow-driving limit and those with jumps only at the boundaries in the rapid-driving limit. We infer that this intermediate jumping behavior originates from the step-like nature of the Fermi-Dirac distribution at low temperatures, where a small change in the energy level causes a large change in the system's occupancy. This result indicates that the simple jump-and-hold paradigm breaks down, and intermediate steps are required to suppress work fluctuations while managing dissipation.

Overall, our findings demonstrate that optimal thermodynamic control in the intermediate-time regime exhibits rich and unique structures that cannot be captured by simple interpolations of the slow and rapid limits. The discontinuous switching of protocol families and the emergence of multi-step jumps provide new physical insights into the interplay between quantum dynamics, environmental coupling, and thermodynamic fluctuations. We believe that our approach and findings pave the way for designing highly efficient and stable quantum thermal machines, and could stimulate further theoretical and experimental investigations into optimal control in non-equilibrium open quantum systems.

\appendix

\section{Methods of minimization of fluctuation and dissipation}
\label{sec:methods}
We consider a quantum system whose Hamiltonian $H(u(t))$ depends on a
control parameter $u(t)$, where
our objective is to minimize the cost functional defined in Eq.~\eqref{eq:cost}.
%In order to evaluate the work variance~\eqref{eq:fluctuation}, we introduce an auxiliary operator $Y(t)$ defined by
However, the work variance in Eq.~\eqref{eq:fluctuation} is nonlocal in time:
the integrand at time $t_1$ depends on the state at earlier times $t_2<t_1$
through the two-time propagator $\overleftarrow{\mathcal P}(t_1,t_2)$.
Therefore, $\sigma_w^2$ cannot be directly incorporated as a running cost in the
standard GRAPE formulation, where the objective functional is assumed to be of the
time-local form~\eqref{eq:PMPcost} as in Appendix~\ref{sec:GRAPE}.
To cast the variance minimization into this framework, we introduce an auxiliary
operator $Y(t)$ defined by Eq.~\eqref{eq:Y_def}.
This construction is formally equivalent to the auxiliary-operator representation used in
Ref.~\cite{erdman_PRR_2023} for work/power fluctuations in cyclic quantum heat engines,
while here we employ it in the TPM setting for general finite-time driving with fixed boundary conditions and gradient-based optimal control.
%To cast the variance minimization into this framework, \kuro{we use} an auxiliary operator $Y(t)$ defined by Eq.~\eqref{eq:Y_def}\kuro{\cite{erdman_PRR_2023}}.
\begin{comment}
\begin{equation}
\label{eq:Y_def}
Y(t) := \int_0^{t}
\overleftarrow{\mathcal P}(t,\tau)\bigl[S_{\rho(\tau)}(\dot H(\tau))\bigr]\,d\tau.
\end{equation}
\end{comment}
%\hase{カンマをピリオドに}
%As shown in Appendix~\ref{sec:Y(t)}, 
%\kuro{appendix F を　ここに吸収する}

By definition~\eqref{eq:propagator},\eqref{eq:Y_def},and the Leibniz integral rule,
\begin{comment}
\begin{align}
\partial_{t_1}\overleftarrow{\mathcal P}(t_1,t_2)
&= \mathcal L_{t_1}\,\overleftarrow{\mathcal P}(t_1,t_2),\qquad
\overleftarrow{\mathcal P}(t_2,t_2) = \mathcal I .
\label{eq:P_property}
\end{align}
\end{comment}

\begin{align*}
    &\frac{d}{dt} Y(t) \\
    &=\frac{d}{dt}\int_{0}^{t}\overleftarrow{\mathcal{P}}(t,\tau)\bigl[S_{\rho(\tau)}(\dot{H}(\tau))\bigr]\,d\tau\\
    &=\overleftarrow{\mathcal{P}}(t,t)S_{\rho(t)}(\dot{H}(t))+\int_{0}^{t}d\tau\frac{\partial}{\partial t}\overleftarrow{\mathcal{P}}(t,\tau)\bigl[S_{\rho(\tau)}(\dot{H}(\tau))\bigr]\\\
    &=S_{\rho(t)}(\dot{H}(t))+\int_{0}^{t}d\tau\,\mathcal{L}_{t}\left[\overleftarrow{\mathcal{P}}(t,\tau)\bigl[S_{\rho(\tau)}(\dot{H}(\tau))\bigr]\right]\\
    &=S_{\rho(t)}(\dot{H}(t))+\mathcal{L}_{t}\left[\int_{0}^{t}d\tau \overleftarrow{\mathcal{P}}(t,\tau)\bigl[S_{\rho(\tau)}(\dot{H}(\tau))\bigr]\right]\\
    &=S_{\rho(t)}(\dot{H}(t))+\mathcal{L}_{t}\left[Y(t)\right]
\end{align*}

$Y(t)$ obeys the following time-local equation of motion\cite{erdman_PRR_2023}:
\begin{equation}
\label{eq:Y_ode}
\dot Y(t) = \mathcal L_t\bigl[Y(t)\bigr] + S_{\rho(t)}(\dot H(t)),\qquad
Y(0)=0 .
\end{equation}
%, \hase{カンマ消す}and 
Substituting~\eqref{eq:Y_def} into~\eqref{eq:fluctuation}
yields the equivalent single-integral representation
\begin{equation}
\label{eq:sigma_single_integral}
\sigma_w^2 = 2\int_0^T \tr\bigl(\dot H(t)\,Y(t)\bigr)\,dt .
\end{equation}
Consequently, by augmenting the state with $Y(t)$, the variance contribution becomes
a running cost depending only on instantaneous variables, which makes the GRAPE optimization applicable.
Additionally, by treating $Y(t)$ as a state variable and storing it at each time step as we do for $\rho(t)$, $Y(t+\Delta t)$ can be updated from $Y(t)$ using Eq.~\eqref{eq:Y_ode}. With this approach, we avoid the direct evaluation of the double time integral in Eq.~\eqref{eq:fluctuation} and can compute it efficiently.

Using Eqs.~\eqref{eq:def_W_sigma}, \eqref{eq:dissipation}, and
\eqref{eq:sigma_single_integral}, the cost functional defined in Eq.~\eqref{eq:cost} can be
cast into the standard form
\begin{equation}
J = \int_0^T L(t)\,dt + \phi(T),
\end{equation}
where the running cost $L(t)$ and the terminal cost $\phi(T)$ are given by
\begin{align}
L(t) &= (1-\alpha)\,\tr\bigl(\dot H(t)\,\rho(t)\bigr)
     + \alpha\beta\,\tr\bigl(\dot H(t)\,Y(t)\bigr), \\
\phi(T) &= - (1-\alpha)\,\Delta F
          + \frac{\kappa}{2}\bigl(u(T)-u_T\bigr)^2 .
\end{align}

Here we used $W_{\mathrm{diss}} = \langle W\rangle - \Delta F$ and
$\langle W\rangle = \int_0^T \tr(\dot H(t)\rho(t))\,dt$, so that the
contribution $-\,(1-\alpha)\Delta F$ appears in the terminal cost.
%\kuro{GRAPE algorithm は コスト関数が操作パラメータの関数であることを要求する。今、操作パラメータをuとすると、コスト関数に udot が出てきてしまうので操作パラメータを$v(t)=\dot{u}(t)$に取り直しておく。その上で、uは$\dot{u}(t)=v(t)$ と言う時間発展方程式に従う状態変数として扱う。}
The GRAPE algorithm requires the cost functional to be expressed explicitly as a functional of the control parameter.
If we choose $u(t)$ as the control, the running cost depends on $\dot u(t)$ through $\dot H(t)$, and the objective is not in the standard form with respect to $u(t)$.
We therefore reparameterize the control by introducing
$v(t):=\dot u(t)$,
and treat $u(t)$ as an additional state variable governed by the kinematic equation $\dot{u}(t)=v(t)$.
We further assume that the Hamiltonian depends on $u$ only through a smooth
operator-valued function $G(u)$ such that
\begin{equation}
\label{eq:G}
\dot H(t) = \frac{\partial H}{\partial u}(u(t))\,\dot u(t)
         = G(u(t))\,v(t).
\end{equation}
Then $S_{\rho(t)}(\dot H(t))$ is linear in $v(t)$ and can be written as
\[
S_{\rho(t)}(\dot H(t))
= v(t)\,S_{\rho(t)}\bigl(G(u(t))\bigr).
\]

%\hase{vectorizationにはcolumn stackとrow stack流派があるので、明記した方がよい。後で具体的な行列が出てくるので。}
%\hase{後ろを見ると、column stackとかではなくて、非対角は実部虚部でとっているんですね。そこらへんも説明が必要かと。この段階では、どの形でも記述は可能だと思うけど、それならそうと書く。}
Upon an arbitrary vectorization~\cite{vectorization}, we denote by $\bm{x}(t)$ and
$\bm{y}(t)$ the vectorized forms of $\rho(t)$ and $Y(t)$, respectively, and by
$A(t)$ the matrix representation of the Lindblad generator $\mathcal L_t$,
such that $\partial_t \bm{x}(t) = A(t)\bm{x}(t)$.
We also introduce the vector
$\bm{s}\bigl(\bm{x}(t),u(t)\bigr)$ corresponding to
$S_{\rho(t)}\bigl(G(u(t))\bigr)$.

With the introduction of the auxiliary operator $Y(t)$, together with the vectorization and the reparametrization of the control in terms of $v(t)=\dot u(t)$, the following GRAPE algorithm becomes applicable.
\begin{enumerate}[label=\arabic*., leftmargin=*, itemsep=2pt]
  \item For a given trial control $v(t)$, propagate the state variables
        $u(t)$, $\bm{y}(t)$, and $\bm{x}(t)$ forward in time using
        Eqs.~\eqref{eq:u}–\eqref{eq:x}.
  \item Set the terminal values of the adjoint variables according to Eqs.~\eqref{eq:terminal1}-\eqref{eq:terminal3} at $t=T$.
  \item Propagate the adjoint variables backward in time using the adjoint
        equations and the Pontryagin Hamiltonian $H_{\mathrm{pmp}}(t)$ in Eqs.~\eqref{eq:Hpmp} and \eqref{eq:backward1}-\eqref{eq:backward3}.
  \item Compute $\delta v(t) = \partial H_{\mathrm{pmp}}/\partial v(t)$ and update
        the control according to
        $v(t) \rightarrow v(t) - \eta\,\delta v(t)$,
        with a suitable step size $\eta>0$.
  \item Repeat steps 1–4 until convergence of the cost $J$.
\end{enumerate}

In step 1,
the dynamics of the state variables is then expressed as
\begin{align}
\dot{u}(t)      &= v(t),                                  \label{eq:u}\\
\dot{\bm{y}}(t) 
&= A(u(t))\,\bm{y}(t)+ v(t)\,\bm{s}\bigl(\bm{x}(t),u(t)\bigr), \label{eq:y}\\
\dot{\bm{x}}(t) &= A(u(t))\,\bm{x}(t).                        \label{eq:x}
\end{align}

Next, in step 2, we associate adjoint variables
$p(t)$, $\Lambda(t)$, and $\Pi(t)$ with $u(t)$, $\bm{y}(t)$, and
$\bm{x}(t)$, respectively.
The terminal values of the adjoint variables are determined from the terminal
cost,
\begin{align}
\bm{\Pi}(T)
&= \frac{\partial \phi(T)}{\partial \bm{x}(T)}
 = \bm{0}, \label{eq:terminal1}\\
\bm{\Lambda}(T)
&= \frac{\partial \phi(T)}{\partial \bm{y}(T)}
 = \bm{0}, \label{eq:terminal2}\\[1ex]
p(T)
&= \frac{\partial \phi(T)}{\partial u(T)} \notag\\
&= -(1-\alpha)\,\frac{\partial \Delta F}{\partial u(T)}
   + \kappa\bigl(u(T)-u_T\bigr).\label{eq:terminal3}
\end{align}

Then in step 3,
the Pontryagin Hamiltonian is defined as
\begin{align}
\label{eq:Hpmp}
H_{\mathrm{pmp}}(t)
=& L(t)
+ p(t)\,v(t)
+ \Lambda(t)^\top\bigl[A(u(t))\bm{y}(t)
+ v(t)\,\bm{s}\bigl(\bm{x}(t),u(t)\bigr)\bigr] \nonumber \\
&+ \Pi(t)^\top A(u(t))\bm{x}(t).
\end{align}

Their time evolution is obtained from
\begin{align}
\dot{\bm{\Pi}}(t) &= -\frac{\partial H_{\mathrm{pmp}}}{\partial \bm{x}}(t) \notag\\
    &= -A\bigl(u(t)\bigr)^{\top}\bm{\Pi}(t)
       - (1-\alpha)\, v(t)\, \bm{g}\bigl(u(t)\bigr) \notag\\
    &\quad - v(t)\, J\bigl(\bm{x}(t),u(t)\bigr)^{\top}\bm{\Lambda}(t), \label{eq:backward1}\\[2ex]
\dot{\bm{\Lambda}}(t) &= -\frac{\partial H_{\mathrm{pmp}}}{\partial \bm{y}}(t) \notag\\
    &= -A\bigl(u(t)\bigr)^{\top}\bm{\Lambda}(t)
       - \alpha\beta\, v(t)\, \bm{g}\bigl(u(t)\bigr), \label{eq:backward2}\\[2ex]
\dot{p}(t) &= -\frac{\partial H_{\mathrm{pmp}}}{\partial u}(t) \notag\\
&= - \bm{\Lambda}(t)^\top\partial_u A\bigl(u(t)\bigr)\,\bm{y}(t)
   - \bm{\Pi}(t)^\top    \partial_u A\bigl(u(t)\bigr)\,\bm{x}(t) \notag\\
&\quad - v(t)\Bigl[(1-\alpha)\,\partial_u\bm{g}\bigl(u(t)\bigr)^\top \bm{x}(t)
                  +\alpha\beta\,\partial_u\bm{g}\bigl(u(t)\bigr)^\top \bm{y}(t)\Bigr]\notag\\
   &- v(t)\,\bm{\Lambda}(t)^\top \partial_u \bm{s}\bigl(\bm{x}(t),u(t)\bigr).
\label{eq:backward3}
\end{align}

For effective calculation, $\partial_u A(u(t))$ should be computed analytically in advance.

Finally in step 4,
$\bm{g}(u)$ denotes the vectorized form of the operator $G(u)$, satisfying $\tr[G(u)\rho] = \bm{g}(u)^{\top}\bm{x}$, and the Jacobian is defined as $J(\bm{x},u) := \partial \bm{s}(\bm{x},u) / \partial \bm{x}$. Then, the gradient with respect to the control $v(t)$ can be calculated as
\begin{align}
\delta v(t) &:= \frac{\partial H_{\mathrm{pmp}}}{\partial v}(t) \notag\\
     &= (1-\alpha)\,\bm{g}(u(t))^{\top}\bm{x}(t)
      + \alpha\beta\,\bm{g}(u(t))^{\top}\bm{y}(t) \notag\\
     &\quad + p(t)
      + \bm{\Lambda}(t)^{\top} \bm{s}\bigl(\bm{x}(t),u(t)\bigr) .
\end{align}

\section{optimization of spin-boson model}
\label{appendix:spin-boson}

For the spin-boson model, we select vectorization as 
$\bm{x}=[\rho_{gg},\rho_{ee},\Re\rho_{eg},\Im\rho_{eg}]^\top$ and
$\bm{y}=[Y_{gg},Y_{ee},\Re Y_{eg},\Im Y_{eg}]^\top$.
In this basis, the Lindblad generator takes the matrix form and can be written as $\dot{\bm{x}}=A(u(t))\bm{x}$,
where
\begin{equation}
\label{eq:matrix_A}
A\bigl(u(t)\bigr)=C\bigl(u(t)\bigr)+R\bigl(-\theta(t)\bigr)\,L\bigl(E(t)\bigr)\,R\bigl(\theta(t)\bigr).
\end{equation}
Here \(C(u)\) is the contribution of the commutator term \(-i[H(t),\rho(t)]\), given by
\begin{align}
C(u)=
\begin{pmatrix}
0&0&0&2\Delta\\
0&0&0&-2\Delta\\
0&0&0&2u\\
-\Delta&\Delta&-2u&0
\end{pmatrix}.
\label{eq:Cmatrix}
\end{align}
The matrix \(L(E)\) denotes the generator of the jump (dissipative) part in the instantaneous
energy eigenbasis, %\(\tilde\rho=U(\theta)\rho U(\theta)^\dagger\), 
and takes the form
\begin{align}
L(E)&=
\begin{pmatrix}
-\gamma_\uparrow(E)&\gamma_\downarrow(E)&0&0\\
\gamma_\uparrow(E)&-\gamma_\downarrow(E)&0&0\\
0&0&-\dfrac{\gamma_\Sigma(E)}{2}&0\\
0&0&0&-\dfrac{\gamma_\Sigma(E)}{2}
\end{pmatrix},
\notag\\
\gamma_\Sigma(E)&:=\gamma_\uparrow(E)+\gamma_\downarrow(E).
\label{eq:Lmatrix}
\end{align}

The matrix \(R(\theta)\) represents the basis rotation $\tilde\rho=\exp(-i\sigma_y \theta/2)\rho \exp(i\sigma_y \theta/2)$
in the above real vectorization \(x\), and is given by
\begin{align}
R(\theta)=
\begin{pmatrix}
\dfrac{1+\cos\theta}{2}&\dfrac{1-\cos\theta}{2}&\sin\theta&0\\[4pt]
\dfrac{1-\cos\theta}{2}&\dfrac{1+\cos\theta}{2}&-\sin\theta&0\\[4pt]
-\dfrac{\sin\theta}{2}&\dfrac{\sin\theta}{2}&\cos\theta&0\\[4pt]
0&0&0&1
\end{pmatrix},\\
\qquad
R(-\theta)=R(\theta)\big|_{\sin\theta\mapsto-\sin\theta}.
\label{eq:Rmatrix}
\end{align}

In this basis, the vectorized control operator $\bm{g}(u)$, defined by $\tr[G(u)\rho]=\bm{g}(u)^{\top}\bm{x}$, is constant and given by $\bm{g}= [-1,1,0,0]^\top$
which corresponds to $G(u)=\sigma_z$.
%with diagonal elements $-1$ and $+1$ corresponding to the states $|g\rangle$ and $|e\rangle$, respectively. 
In this setting, since $\dot H(t)=v(t)\sigma_z$, the anticommutator term in Eq.~\eqref{eq:fluctuation} satisfies
\begin{equation}
S_\rho(\dot H(t)) = v(t)\,S_\rho(\sigma_z),
\qquad
S_{\rho}(\sigma_z) :=\frac{1}{2}\{ \rho, \Delta_\rho \sigma_z \}_+ .
\end{equation}
Since $\rho$ is a density operator, we use  $\rho_{gg}+\rho_{ee}=1$ to simplify the explicit expression of $S_\rho(\sigma_z)$.
Upon vectorization, $\operatorname{vec}\left(S_\rho(\sigma_z)\right)=\bm{s}(\bm{x})$ takes the form
\begin{equation}
\label{eq:s_vector}
\bm{s}(\bm{x}) = 
\begin{pmatrix}
-2x_1 x_2 \\
2x_1 x_2 \\
(x_1 - x_2) x_3 \\
(x_1 - x_2) x_4
\end{pmatrix},
\end{equation}
where we used the state vector components $x_1=\rho_{gg}$, $x_2=\rho_{ee}$, $x_3=\Re\rho_{eg}$, and $x_4=\Im\rho_{eg}$.
Consequently, the Jacobian $J(\bm{x}) = \partial \bm{s}/\partial \bm{x}$ is given by
\begin{equation}
\label{eq:j_matrix}
J(\bm{x}) = 
\begin{pmatrix}
-2x_2 & -2x_1 & 0 & 0 \\
2x_2 & 2x_1 & 0 & 0 \\
x_3 & -x_3 & x_1-x_2 & 0 \\
x_4 & -x_4 & 0 & x_1-x_2
\end{pmatrix}.
\end{equation}

We precompute $\partial_u A\left(u(t)\right)$ analytically to evaluate Eq.~\eqref{eq:backward3}, as detailed in Appendix~\ref{sec:differential}.
We also evaluate Eq.~\eqref{eq:terminal3} numerically with
$E_T=\sqrt{u(T)^2+\Delta^2}$, which yields
\begin{equation}
p(T)=(1-\alpha)\tanh(\beta E_T)\frac{u(T)}{E_T}+\kappa\bigl(u(T)-u_T\bigr).
\end{equation}
Using the component-wise representations introduced above, we implement the numerical procedure described in the Methods section.
\begin{comment}
The numerical settings used in the calculations are as follows.
We discretize the time interval $[0,T]$ into $N=10{,}000$ uniform steps with $\Delta t=T/N$ (i.e., $N+1$ grid points).
The control is parameterized by $v(t)=\dot u(t)$ and represented as a piecewise-constant function on each interval $[t_k,t_{k+1})$, while $u(t)$ is obtained by forward integration,
$u_{k+1}=u_k+\Delta t\,v_k$,
with the fixed initial condition $u(0)=u_0$.
We use an explicit Euler scheme for the forward propagation of $x(t)$ and $y(t)$ and for the backward propagation of the adjoint variables.
We initialize the optimization with a linear ramp connecting $u_0$ and $u_T$ (i.e., constant $v=(u_T-u_0)/T$), and update $v$ by gradient descent with learning rate $\eta=0.01$ for up to $N_{\mathrm{iter}}=10^6$ iterations.
The terminal condition is enforced via the quadratic penalty $\frac{\kappa}{2}\bigl(u(T)-u_T\bigr)^2$ with $\kappa=10.0$.
For numerical stability, we allow box bounds $u(t)\in[-8,8]$ and $v(t)\in[-100,100]$ (implemented via projection after each update); for all results reported here these bounds are inactive, i.e., $u(t)$ and $v(t)$ remain well within the prescribed ranges.
Unless otherwise stated, we set $\beta=1$; for the quantum-dot model we also set $\Gamma=1$.
\kuro{T=10 で成立しないから要修正}
As specific parameter settings, we examine \romannumeral 1)~$T=1,\Delta=0$, \romannumeral 2)~$T=10,\Delta=0$, and \romannumeral 3)~$T=1,\Delta=1$.
\end{comment}

The numerical settings used in the calculations are as follows.
We discretize the time interval $[0,T]$ into $N$ uniform steps with $\Delta t=T/N$.
In practice, we choose $N$ so that the time step is kept fixed across different protocol durations; specifically, we use $N=1{,}000$ for $T=1$ and $N=10{,}000$ for $T=10$.
The control is parameterized by $v(t)=\dot u(t)$ and represented as a piecewise-constant function on each interval $[t_k,t_{k+1})$, while $u(t)$ is obtained by forward integration,
$u_{k+1}=u_k+\Delta t\,v_k$,
with the fixed initial condition $u(0)=u_0$.
We use an explicit Euler scheme for the forward propagation of $x(t)$ and $y(t)$ and for the backward propagation of the adjoint variables.
We initialize the optimization with a linear ramp connecting $u_0$ and $u_T$ (i.e., constant $v=(u_T-u_0)/T$), and update $v$ by gradient descent with learning rate $\eta$ for up to $N_{\mathrm{iter}}=10^6$ iterations.
We take $\eta=0.01$ for $T=1$ and $\eta=0.001$ for $T=10$.
The terminal condition is enforced via the quadratic penalty $\frac{\kappa}{2}\bigl(u(T)-u_T\bigr)^2$ with $\kappa=10.0$.
For numerical stability, we allow box bounds $u(t)\in[-8,8]$ and $v(t)\in[-100,100]$ (implemented via projection after each update); for all results reported here these bounds are inactive, i.e., $u(t)$ and $v(t)$ remain well within the prescribed ranges.
Unless otherwise stated, we set $\beta=1$.

\section{rapid drive approximation}
\label{sec:rapid drive}
Under the rapid drive approximation \cite{rapid_drive}, which 
%assumes initial and final jump protocol and constant between them, 
assumes an initial and a final jump, with a constant segment between them,
excess work and fluctuations in the setting of section~\ref{subsec:spin-boson}  can be written as
\begin{align}
\label{app:eq:Wex-trace}
 W_{\mathrm{diss}}
&= \beta^{-1}\,S\big(\pi(0)\,\Vert\,\pi(u_T)\big) \\ \notag
&+\int_{0}^{T}dt\,\big[u_T-u(t)\big]R\big(u(t)\big),
\end{align}
%式31
\begin{align}
\label{app:eq:var-trace-noadj}
\sigma_w^{2}
=& \beta^{-2}\,V\big(\pi(0)\,\Vert\,\pi(u_T)\big)\\ \notag
&+\int_{0}^{T}dt\Big\{
\big[u_T-u(t)\big]^2\,G\big(u(t)\big) \\ \notag
&+\big[u_T-u(t)\big]B\big(u(t)\big)u(t)
\Big\},
\end{align}
Here $S$ and $V$ are relative entropy and relative entropy variance respectively.

\begin{align}
S(\rho_1 || \rho_2) &=  \tr[\rho_1 \log \rho_1] - \tr[\rho_1 \log \rho_2]   \\
V(\rho_1 || \rho_2) & = \tr[\rho_1 (\log \rho_1 - \log \rho_2)^2 ] -S^2(\rho_1 || \rho_2) 
\end{align}
$R(u),G(u)$ and $B(u)$ are defined as 
\begin{align}
\label{app:eq:R-def-noadj}
R(u)&:=\tr\big[\sz\,\mathcal{L}_{u}\big(\pi(0)\big)\big],
\\[4pt]
\label{app:eq:G-def-noadj}
G(u)&:=\frac{1}{2}\,\tr\Big[\big\{\sz,\sz\big\}_+\,\mathcal{L}_{u}\big(\pi(0)\big)\Big],
\\[4pt]
\label{app:eq:B-def-noadj}
B(u)&:=\tr\Big[\sz\,\mathcal{L}_{u}\big(\big\{\sz,\pi(0)\big\}_+\big)\Big],
\end{align}
and $\pi(u):=e^{-\beta u\,\sz}/\bigl(2\cosh(\beta u)\bigr)$, $\{\cdot,\cdot\}_+$ is the anticommutator, and $\gamma(u)=u^3$.
The dissipation-minimizing jump $\zeta$ satisfies
\begin{equation}
\left.\frac{d}{du}\Big[(u_T-u)\,R(u)\Big]\right|_{u=\zeta}=0,
\end{equation}
while the fluctuation-minimizing jump $\Lambda$ satisfies
\begin{equation}
\left.\frac{d}{du}\Big[(u_T-u)^{2}\,G(u)+(u_T-u)\,B(u)\,u\Big]\right|_{u=\Lambda}=0.
\end{equation}
%式 30,31

By applying these equations to our settings,
\begin{align}
\label{app:eq:R-closed}
R(u)&= -\,\gamma(u),
\\
\label{app:eq:G-closed}
G(u)&= 0,
\\
\label{app:eq:B-closed}
%B(u)&= -\,2\,\gamma(u)\,\coth(\beta u), \\
B(u)&= -2\gamma(u)\coth(\beta u),
\end{align}
with $\gamma(u)=u^3$. Then we obtain dissipation minimizing jump $\zeta$ and fluctuation minimizing jump $\Lambda$ satisfy :
\begin{align}
\label{app:eq:zeta-result}
\zeta&=\frac{3}{4}\,u_T,
\\
\label{app:eq:Lambda-result}
\bigl(5\Lambda-4u_T\bigr)\coth(\beta \Lambda)
&=\beta\,\Lambda\,(\Lambda-u_T)\,\mathrm{csch}^{2}(\beta \Lambda).
\end{align}
%\kuro{A15の計算間違いを修正}
at $u_T=1, \beta =1$, optimal protocol jumps at $\zeta=0.6$ and $\Lambda=0.77$ respectively.

\section{GRAPE algorithm}
\label{sec:GRAPE}
We use the GRAPE algorithm to optimize the protocol in this paper.
We then summarize the algorithm from \cite{PMP}.
As a problem setting, the dynamics of the system is described as 
\begin{equation}
\dot{q}(t)= f\bigl(q(t),u(t)\bigr),
\end{equation}
where $q(t)$ is the state vector and $u(t)$ is the control parameter,
and the cost function to be minimized during a fixed time duration $[0,T]$ is described as
\begin{equation}
\label{eq:PMPcost}
J=\int_0^T f^{0}\bigl(q(t),u(t)\bigr)\,dt + d\bigl(T,q(T)\bigr) .
\end{equation}

Pontryagin Hamiltonian is defined as
\begin{equation}
\mathcal{H}(q,p,u)=p^{\top} f(q,u)+ f^{0}(q,u),
\end{equation}
where $p(t)$ is the costate variable.
%Then the optimal protocol $\bar{u}(t)$ satisfies the necessary conditions of Pontryagin’s maximum principle (PMP).
If $u(t)$ is optimal, then there exists an adjoint state $p(t)$ such that $(q(t),p(t),u(t))$ satisfies the necessary conditions of Pontryagin’s principle:
\begin{align}
\dot{q}(t)&= \frac{\partial \mathcal{H}}{\partial p}\bigl(q(t),p(t),u(t)\bigr) = f\bigl(q(t),u(t)\bigr), \\
\dot{p}(t) &=- \left(\frac{\partial \mathcal{H}}{\partial q}\bigl(q(t),p(t),u(t)\bigr)\right)^\top, \\
p(T) &= \left.\left(\frac{\partial d(T,q)}{\partial q}\right)^\top\right|_{q=q(T)}, \\
\frac{\partial \mathcal{H}}{\partial u}&\bigl(q(t),p(t),u(t)\bigr) = 0 .
\end{align}

In order to satisfy these equations,
the GRAPE algorithm is as follows.

\begin{enumerate}[label=\arabic*., leftmargin=*, itemsep=2pt]
  \item For a given trial control $u(t)$, propagate the state variable
        $q(t)$ forward in time by solving
        \begin{equation}
          \dot{q}(t)=f\bigl(q(t),u(t)\bigr),\qquad q(0)=q_0 .
        \end{equation}
  \item Set the terminal value of the adjoint variable at $t=T$ as
        \begin{equation}
          p(T)= \left.\left(\frac{\partial d(T,q)}{\partial q}\right)^\top\right|_{q=q(T)} .
        \end{equation}
  \item Propagate the adjoint variable $p(t)$ backward in time using the adjoint
        equation
        \begin{equation}
          \dot{p}(t)=-\left(\frac{\partial \mathcal{H}}{\partial q}\bigl(q(t),p(t),u(t)\bigr)\right)^{\top}.
        \end{equation}
  \item Compute the gradient
        \[
          g(t)=\frac{\partial \mathcal{H}}{\partial u}\bigl(q(t),p(t),u(t)\bigr),
        \]
        and update the control according to
        \[
          u(t)\rightarrow u(t)-\eta\, g(t),
        \]
        with a suitable step size $\eta>0$.
  \item Repeat steps 1--4 until convergence of the cost.
\end{enumerate}

\section{calculation of $\partial_u A(u)$}
We analytically calculate $\partial_u A(u)$ beforehand for the GRAPE algorithm.
\label{sec:differential}
\begin{align}
\partial_u A(u)
&=\partial_u C(u)+\partial_u\Bigl(R(-\theta)\,L(E)\,R(\theta)\Bigr) \notag\\
&=\partial_u C(u)
+\bigl(\partial_u R(-\theta)\bigr)L(E)R(\theta) \notag\\
&\quad +R(-\theta)\bigl(\partial_u L(E)\bigr)R(\theta)
+R(-\theta)L(E)\bigl(\partial_u R(\theta)\bigr).
\label{eq:dAdu_general}
\end{align}
\begin{comment}
Here \(E=\sqrt{u^2+\Delta^2}\) and the rotation angle is defined by
\begin{equation}
\cos\theta=\frac{u}{E},\qquad \sin\theta=\frac{\Delta}{E}.
\label{eq:theta_def}
\end{equation}
Therefore,
\end{comment}
As defined in Eqs.\eqref{eq:theta_def},
\begin{align}
\frac{dE}{du}=\frac{u}{E},\qquad
\frac{d}{du}\cos\theta=\frac{\Delta^2}{E^3},\qquad
\frac{d}{du}\sin\theta=-\frac{u\Delta}{E^3}.
\label{eq:dE_dc_ds}
\end{align}

The coherent contribution yields
\begin{align}
\partial_u C(u)=
\begin{pmatrix}
0&0&0&0\\
0&0&0&0\\
0&0&0&2\\
0&0&-2&0
\end{pmatrix}.
\label{eq:dCdu}
\end{align}

Since \(L(E)\) depends on \(u\) only through \(E\), the chain rule gives
\begin{align}
\partial_u L(E)
&=\frac{dE}{du}\,\partial_E L(E) \notag\\
&=\frac{u}{E}\,
\begin{pmatrix}
-\gamma_\uparrow'(E)&\gamma_\downarrow'(E)&0&0\\
\gamma_\uparrow'(E)&-\gamma_\downarrow'(E)&0&0\\
0&0&-\dfrac{\gamma_\Sigma'(E)}{2}&0\\
0&0&0&-\dfrac{\gamma_\Sigma'(E)}{2}
\end{pmatrix},
\notag\\
\gamma_\Sigma'(E)&=\gamma_\uparrow'(E)+\gamma_\downarrow'(E),
\label{eq:dLdu}
\end{align}
%where the prime denotes differentiation with respect to \(E\).
where the prime denotes differentiation with respect to $E$. In particular, with Eqs.~\eqref{eq:sb_rates},
we have
\begin{align}
P'(E) &= -2\beta\,P(E)\bigl(P(E)+1\bigr),\\
\gamma'_\uparrow(E) &= \gamma'(E)P(E)+\gamma(E)P'(E),\\
\gamma'_\downarrow(E) &= \gamma'(E)\bigl(P(E)+1\bigr)+\gamma(E)P'(E),
\end{align}
with $\gamma'(E)=3E^2$.

Moreover, \(R(\theta)\) depends on \(u\) through \(c=\cos\theta\) and \(s=\sin\theta\) in
\eqref{eq:Rmatrix}, so that
\begin{align}
\partial_u R(\theta)
=\frac{dc}{du}\,\partial_c R + \frac{ds}{du}\,\partial_s R,
\label{eq:dRdu_chain}
\end{align}
with
\begin{align}
\partial_c R=
\begin{pmatrix}
\frac12&-\frac12&0&0\\
-\frac12&\frac12&0&0\\
0&0&1&0\\
0&0&0&0
\end{pmatrix},
\qquad
\partial_s R=
\begin{pmatrix}
0&0&1&0\\
0&0&-1&0\\
-\frac12&\frac12&0&0\\
0&0&0&0
\end{pmatrix}.
\label{eq:dRdc_dRds}
\end{align}
Since \(R(-\theta)\) is obtained by the substitution \(s\mapsto -s\), we have
\begin{align}
\partial_u R(-\theta)
=\frac{dc}{du}\,\partial_c R -\frac{ds}{du}\,\partial_s R.
\label{eq:dRminusdu}
\end{align}

Combining \eqref{eq:dAdu_general} with \eqref{eq:dE_dc_ds}--\eqref{eq:dRminusdu} yields an explicit
expression for \(\partial_u A(u)\).

\section{Equivalence between the Fock-space description of a single-level quantum dot and a two-level (pseudospin) representation}
\label{sec:qd_fock_to_tls}
%Feedback stabilization of pure states in quantum transport
\paragraph{Quantum dot in the Fock basis.}
A single-level quantum dot (neglecting spin degeneracy) is naturally described in the fermionic Fock space spanned by the empty and occupied states $\{\ket{0},\ket{1}\}$.
The annihilation and creation operators $a$ and $a^\dagger$ act as
\begin{equation}
a\ket{0}=0,\quad a\ket{1}=\ket{0},\quad
a^\dagger\ket{0}=\ket{1},\quad a^\dagger\ket{1}=0.
\label{eq:fermion_actions}
\end{equation}
In the ordered basis $(\ket{0},\ket{1})$, these operators have the matrix representations
\begin{equation}
\ket{0}=\begin{pmatrix}1\\0\end{pmatrix},\quad
\ket{1}=\begin{pmatrix}0\\1\end{pmatrix},\quad
a=\begin{pmatrix}0&1\\0&0\end{pmatrix},\quad
a^\dagger=\begin{pmatrix}0&0\\1&0\end{pmatrix}.
\label{eq:matrix_rep_a}
\end{equation}
Hence the number operator is
\begin{equation}
n=a^\dagger a=\begin{pmatrix}0&0\\0&1\end{pmatrix}=\ket{1}\bra{1}.
\label{eq:number_op_projector}
\end{equation}
The standard dot Hamiltonian is then
\begin{equation}
H_{\mathrm{QD}}(t)=\varepsilon(t)\,a^\dagger a
=\varepsilon(t)\begin{pmatrix}0&0\\0&1\end{pmatrix}.
\label{eq:H_qd_fock}
\end{equation}

\paragraph{Two-level-system representation.}
For a generic two-level system, we use the basis $\{\ket{g},\ket{e}\}$.
In the ordered basis $(\ket{g},\ket{e})$, we represent the basis vectors as
\begin{equation}
\ket{g}=\begin{pmatrix}1\\0\end{pmatrix},\qquad
\ket{e}=\begin{pmatrix}0\\1\end{pmatrix}.
\label{eq:ge_vectors}
\end{equation}
The Pauli matrix $\sigma_z$ is defined by
\begin{equation}
\sigma_z\ket{g}=-\ket{g},\qquad \sigma_z\ket{e}=+\ket{e},
\label{eq:sigmaz_action}
\end{equation}
and therefore has the matrix representation
\begin{equation}
\sigma_z=\begin{pmatrix}-1&0\\0&1\end{pmatrix}
\quad \text{in the basis }(\ket{g},\ket{e}).
\label{eq:sigmaz_matrix}
\end{equation}
Identifying the Fock states with the two-level basis as
\begin{equation}
\ket{g}\equiv\ket{0},\qquad \ket{e}\equiv\ket{1},
\label{eq:identify_ge_01}
\end{equation}
we may rewrite the projector $\ket{1}\bra{1}$ in terms of $I$ and $\sigma_z$:
\begin{equation}
\ket{1}\bra{1}=\frac{I+\sigma_z}{2}.
\label{eq:proj_in_terms_of_sigmaz}
\end{equation}
Therefore,
\begin{equation}
H_{\mathrm{QD}}(t)=\varepsilon(t)\ket{1}\bra{1}
=\frac{\varepsilon(t)}{2}\sigma_z+\frac{\varepsilon(t)}{2}I.
\label{eq:H_decomposition}
\end{equation}
The term proportional to $I$ produces only an overall energy shift and can be dropped since it does not affect the density-matrix dynamics ($[I,\rho]=0$).
Defining the rescaled control amplitude
\begin{equation}
u(t)=\frac{\varepsilon(t)}{2},
\label{eq:u_eps_relation}
\end{equation}
we obtain the effective two-level (pseudospin) Hamiltonian
\begin{equation}
H(t)=u(t)\,\sigma_z,
\label{eq:H_tls_final}
\end{equation}
which is the form used in the main text.
Thus, after a trivial rescaling of the Hamiltonian amplitude, the Fock-space description of the single-level quantum dot is equivalent to a two-level-system representation, and the same component-wise formalism can be applied.

\begin{acknowledgments}

This work was supported by JSPS KAKENHI Grant Numbers JP23K24915 and JP24K03008.

\end{acknowledgments}

\end{document}